\documentclass[english,Conference]{IEEEtran}
\usepackage[T1]{fontenc}
\usepackage[latin9]{inputenc}
\usepackage{verbatim}
\usepackage{units}
\usepackage{amssymb,amsmath,amsthm}
\usepackage[english]{babel}
\usepackage{graphicx}
\usepackage{esint}

\usepackage{tikz,pgf,pgfplots}
\usepgflibrary{shapes}
\usetikzlibrary{%
  arrows,%
  shapes.misc,
  shapes.arrows,%
  shapes,%
  chains,%
  matrix,%
  positioning,
  scopes,%
  decorations.pathmorphing,
  shadows,%
  backgrounds,%
  fit,%
  petri%
}

\newtheorem{lemma}{Lemma}
\newtheorem{theorem}{Theorem}
\newtheorem{remark}{Remark}
\newtheorem{proposition}{Proposition}
\newtheorem{corollary}{Corollary}

\newtheorem{assumption}{Assumption}

\begin{document}
\global\long\def\good{\texttt{g}}

\global\long\def\bad{\texttt{b}}

\global\long\def\type{\mathcal{T}}
\title{On The Performance of Random Block Codes\\
over Finite-State Fading Channels}
\vspace{-1mm}
\author{\IEEEauthorblockN{Fatemeh Hamidi-Sepehr,
Jean-Francois Chamberland,
Henry Pfister}\\
\IEEEauthorblockA{Department of Electrical and Computer Engineering, Texas A{\&}M University}
}

\maketitle
\thispagestyle{empty}
\pagestyle{empty}
\vspace{-5mm}
\begin{abstract}
As the mobile application landscape expands, wireless networks are tasked with supporting multiple connection profiles, including real-time communications and delay-sensitive traffic.
Among many ensuing engineering challenges is the need to better understand the fundamental limits of forward error correction in non-asymptotic regimes.
This article seeks to characterize the performance of block codes over finite-state channels with memory.
In particular, classical results from information theory are revisited in the context of channels with rate transitions, and bounds on the probabilities of decoding failure are derived for random codes.
This study offers new insights about the potential impact of channel correlation over time on overall performance.
\end{abstract}
\vspace{-3.2mm}

\section{Introduction}

As preferred mobile devices shift to advanced smartphones and tablet personal computers, the demand for low-latency, high-throughput wireless service increases rapidly.
The shared desire for a heightened user experience, which includes real-time applications and mobile interactive sessions, acts as a motivation for the study of highly efficient communication schemes subject to stringent delay constraints.
An important aspect of delay-sensitive traffic stems from the fact that its intrinsic delivery requirements preclude the use of asymptotically long codewords.
As such, the insights offered by classical information theory are of limited value in this context.

This article focuses on deriving meaningful performance limits for delay-aware systems operating over channels with memory.
The emphasis is put on identifying upper bounds on the probabilities of decoding failure for systems employing short block-lengths.
This is an essential intermediate step in characterizing the queueing behavior of contemporary communication systems, and it forms the primary goal of our inquiry.

A distinguishing feature of our approach is the accent on channels with memory and state-dependent operation.
More specifically, we are interested in regimes where the block length is of the same order or smaller than the channel memory.
Mathematically, we wish to study the scenario where the mixing time of the underlying finite-state channel is similar to the time necessary to transmit a codeword.
This leads to two important phenomena.
First, the state of the channel at the onset of a transmission has a significant impact on the empirical distribution of the states within a codeword transmission cycle.
Second, channel dependencies extend beyond the boundaries of individual codewords.
This is in stark contrast with block-fading models; for instance, in our proposed framework, decoding failure events can be strongly correlated over time.

Computing probabilities of decoding failures for specific communication channels and fixed coding schemes is of fundamental interest.
This topic has received significant attention in the past, with complete solutions in some cases.
This line of work dates back to the early days of information theory~\cite{Fano-1961}.
An approach that has enjoyed significant success, and chiefly popularized by Gallager, consists in deriving exponential error bounds on the behavior of asymptotically long codewords~\cite{Gallager-1968}.
Such bounds have been examined for memoryless channels as well as finite-state channels with memory.
In general, they can become reasonably tight for long yet finite block-lengths.
It is worth mentioning that the subject of error bounds has also appeared in more recent studies, with the advent of new approaches such as dispersion and the uncertainty-focusing bound~\cite{Barg-it02,Sahai-it08,Polyanskiy-it10,Polyanskiy-it11}.

In standard asymptotic frameworks, channel parameters are kept constant while the length of the codeword increases to infinity.
While these approaches lead to mathematically appealing characterizations, they also have the side effect that the resulting bounds on error probability do not depend on the initial or final states of the channel.
This situation can be attributed to the fact that, no matter how slow the mixing time of the underlying channel is, the length of the codeword eventually far exceeds this quantity. Therefore, the initial and final states of the channel become inconsequential.
Unfortunately, this situation diminishes the value of the corresponding results for queueing models.
Often, in practical scenarios, the service requirements imposed on a communication link forces the use of short codewords, with no obvious time-scale separation between the duration of a codeword and the mixing time of the underlying channel.

This reality, together with the increasing popularity of real-time applications on wireless networks, demands a novel approach where the impact of initial conditions are preserved throughout the analysis.
A suitable methodology should be able to capture both the effects of channel memory as well as the impact of the channel state at the onset of a codeword.
An additional benefit of the slow mixing regime is the ability to track dependencies from codeword to codeword, which intrinsically lead to correlation in decoding failure events and can therefore greatly affect the perceived service quality from a queueing perspective.
In this article, we establish the underpinnings of error probability analysis in the rare-transition regime.

The goal of deriving upper bounds on the probability of decoding failure for rare transitions is to characterize overall performance for systems that transmit data using block lengths on the order of the coherence time of their respective channels.
This article addresses the problem of deriving Gallager-type exponential bounds on the probability of decoding failure in the rare-transition regime.
By construction, these bounds necessarily depend on the initial and final states of the channel.
The analysis is conducted for the scenario where channel state information is available at the receiver.
Our results are then compared to the probability of decoding failure obtained for a Gilbert-Elliott channel under a minimum distance decoder and the maximum-likelihood decision rule~\cite{Fano-1961,Gilbert-bell60,Elliott-bell63}.

After computing the exponential upper bound on error probability,
we consider the rare-transition regime, in which the
number of transitions during a block length $N$, decays with $N$.
We apply this condition on the error exponent and analyze the results
for different $N$s. 

\begin{figure}[tb]
\begin{center}
\begin{tikzpicture}
[node distance = 10mm, draw=black, thick, >=stealth',
state/.style={circle, draw=black, inner sep = 0pt, minimum size = 0.9cm}]
\node[state, fill=black!5] (l0) at (0,0) {1};
\node[state, fill=black!20] (l1) at (0,2) {2}
  edge[<-, dashed, bend right=45] node[left] {\small{$\alpha$}} (l0)
  edge[->, dashed, bend left=45] node[right] {\small{$\beta$}} (l0);
\draw [<-, dashed] (l0) to [in=165,out=210,looseness=5] (l0);
\draw [->, dashed] (l1) to [in=150,out=195,looseness=5] (l1);
\node[coordinate] (l0In0) [right=of l0, yshift=-10, label=left:{\small{$0$}}]{};
\node[coordinate] (l0In1) [right=of l0, yshift=10, label=left:{\small{$1$}}]{};
\node[coordinate] (l0Out0) [right= of l0In0, label=right:{\small{$0$}}]{}
  edge[pre, semithick] node[below] {\scriptsize{$1 - \varepsilon_1$}} (l0In0)
  edge[pre, semithick] node[left] {\scriptsize{$\varepsilon_1$}} (l0In1);
\node[coordinate] (l0Out1) [right= of l0In1, label=right:{\small{$1$}}]{}
  edge[pre, semithick] node[above] {\scriptsize{$1 - \varepsilon_1$}} (l0In1)
  edge[pre, semithick] (l0In0);
\node[coordinate] (l1In0) [right=of l1, yshift=-10, label=left:{\small{$0$}}]{};
\node[coordinate] (l1In1) [right=of l1, yshift=10, label=left:{\small{$1$}}]{};
\node[coordinate] (l1Out0) [right= of l1In0, label=right:{\small{$0$}}]{}
  edge[pre, thick] node[below] {\scriptsize{$1 - \varepsilon_2$}} (l1In0)
  edge[pre, thin] node[left] {\scriptsize{$\varepsilon_2$}} (l1In1);
\node[coordinate] (l0Out1) [right= of l1In1, label=right:{\small{$1$}}]{}
  edge[pre, thick] node[above] {\scriptsize{$1 - \varepsilon_2$}} (l1In1)
  edge[pre, thin] (l1In0);
\end{tikzpicture}
\end{center}\vspace{-3mm}

\caption{The Gilbert-Elliott model is the simplest, non-trivial instantiation of a finite-state channel with memory.
State evolution over time forms a Markov chain, and the input-output relationship of this binary channel is governed by a state-dependent crossover probability, as illustrated above.}
\label{figure:GilbertElliott}
\end{figure}
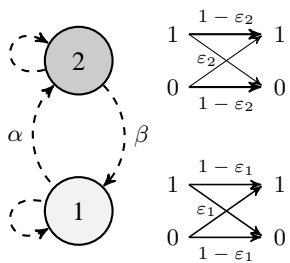
\vspace{-3mm}

\section{Modeling and Exponential Bound}

The models considered in this article belong to the general class of finite-state channels where the state transitions are independent of the input.
Such channels have been used extensively in the information theory literature, and they need very little in way of introduction.
On the other hand, it is important to establish a proper notation.
In this article, the state of the channel at time~$n$ is denoted by $S_n$ and takes value in a finite set.
The corresponding input and output are represented by $X_n$ and $Y_n$, respectively.
Capital letters are used for random variables whereas lower case letters designate elements.
In general, the input-output relationship is governed by Gallager's model, with the conditional probability distribution of the form
\begin{equation*}
\begin{split}
&P \left( y_n, s_n | x_n, s_{n-1} \right) \\
&= \Pr \left( Y_n = y_n, S_n = s_n | X_n = x_n, S_{n-1} = s_{n-1} \right) .
\end{split}
\end{equation*}
In our work, we assume that state transitions are independent of the input so  that the distribution can be factored into two parts,\vspace{-1mm}
\begin{align}
&P \left( y_n, s_n | x_n, s_{n-1} \right)
= P \left( s_n | s_{n-1} \right)
P \left( y_n | x_n, s_n \right) \label{equation:ChannelDecomposition}\\
&=\!\Pr\left(\!S_n \!=\! s_n | S_{n-1} = \!s_{n-1} \right)\!\Pr \left( Y_n \!=\! y_n | X_n = x_n, S_n = s_n \right)\!.\nonumber
\end{align}
The proverbial example for a channel with memory that features this structure is the famed Gilbert-Elliott model, which is governed by a two-state ergodic discrete time Markov chain (DTMC), and is
illustrated in Fig.~\ref{figure:GilbertElliott}.
In this case, the channel evolution forms a Markov chain with transition probability matrix\vspace{-1mm}
\begin{equation*}
\left[ \begin{array}{cc}
P(1|1) & P(1|2) \\
P(2|1) & P(2|2)
\end{array} \right]
= \left[ \begin{array}{cc}
1 - \alpha & \alpha \\
\beta & 1 - \beta
\end{array} \right] .
\end{equation*}
The input-output relation induced by state~$s$ can be written as \vspace{-3mm}
\begin{equation*}
\Pr \left( x_n = y_n | S_n = s \right)
= 1 - \varepsilon_s
\end{equation*}
where $\varepsilon_s$ designates the state-dependent crossover probability.
By convention, we label states so that $\varepsilon_1 \leq \varepsilon_2$. Here we call the first state as the good state, denoted by subscript $\good$ and the second one as the bad state denoted by subscript $\bad$.

A mathematical approach that has proven exceptionally useful in information theory is the use of random codes.
Following this tradition, we adopt a random coding scheme that employs a code ensemble $\mathcal{C}$ with $M = e^{NR}$ elements, where $R$ denotes the rate of the code and $N$ represents the common block length. 
The elements in $\mathcal{C}$ are indexed by $i \in \{ 1, \ldots, M \}$.
Let $Q(x)$ be an arbitrary distribution on the set of possible input symbol.
Throughout, we assume that codewords are selected independently using the corresponding product distribution, $\Pr \left( \underline{X}(i) = \underline{x} \right) = \boldsymbol{Q}_N \left( \underline{x} \right) = \prod_{n=1}^N Q(x_n)$.
A message is transmitted to the destination by selecting one of the codewords.
We wish to upper bound the probability that this codeword is decoded erroneously at the receiver, while also preserving partial state information.

\begin{assumption} \label{assumption:DecompositionRandomCodes}
Communication takes place over a finite-state channel that admits the conditional decomposition of \eqref{equation:ChannelDecomposition}.
Information is transmitted using the coding scheme described above.
Furthermore, the state of the channel is perfectly known at the receiver.
\end{assumption}

For completeness, we reproduce below a celebrated result that we use extensively thereafter; a detailed proof can be found in the associated reference.

\begin{theorem}[Section~5.6, \cite{Gallager-1968}] \label{theorem:GeneralErrorExponent}
Let $P_N (\underline{y} | \underline{x})$ be the transition probability assignment for sequences of length $N \geq 1$ on a discrete channel.
Suppose that maximum-likelihood decoding is employed.
Then, the average probability of decoding error over this ensemble of codes is bounded, for any choice of $\rho$, $0 \leq \rho \leq 1$, by
\begin{equation} \label{equation:GeneralErrorExponent}
P_{\mathrm{e}} \leq (M-1)^{\rho}
\sum_{\underline{y}}
\left[ \sum_{\underline{x}}
\boldsymbol{Q}_N  (\underline{x})
P_N \left( \underline{y} | \underline{x} \right)^{\frac{1}{1+\rho}} \right]^{1 + \rho} .
\end{equation}
\end{theorem}

This theorem is quite general; it applies to channels with memory and, in particular, slow-mixing channels.
$P_N (\cdot|\cdot)$ is a generic conditional probability distribution that can represent the probability distribution induced by a specific channel realization, for instance.
We are now ready to present our first pertinent result.

Consider a channel realization that leads to the sequence $\underline{s}_N = (s_1, \ldots, s_{N})$.
Moreover, let $T = \mathcal{T} (\underline{s}_n)$ denote the empirical distribution of this sequence.

\begin{proposition} \label{proposition:UpperBound}
Suppose Assumption~\ref{assumption:DecompositionRandomCodes} holds.
For any $\rho \in [0,1]$, the probability of decoding failure at the destination, conditioned on $\underline{S}_N = \underline{s}_N$, is bounded by
\begin{equation*}
P_{\mathrm{e}|\underline{s}_N}
\leq \exp \left( - N \left( E_{0,N} (\rho, \boldsymbol{Q}_N, \underline{s}_N) - \rho R \right) \right)
\end{equation*}
where
\begin{equation*}
\begin{split}
&E_{0,N} (\rho, \boldsymbol{Q}_N, \underline{s}_N) \\
&= - \frac{1}{N} \ln \prod_{n=1}^N \sum_{y_n} \left[
\sum_{x_n} Q(x_n) P \left( y_n | x_n, s_n \right)^{\frac{1}{1+\rho}} \right]^{1+\rho} .
\end{split}
\end{equation*}
\end{proposition}
\vspace{-1mm}
\begin{IEEEproof}
Applying Theorem~\ref{theorem:GeneralErrorExponent} to this specific scenario and following the same argument as in \cite[Section~5.5]{Gallager-1968}, we get
\begin{equation*}
\begin{split}
&\sum_{\underline{y}} \left[ \sum_{\underline{x}}
\boldsymbol{Q}_N  (\underline{x})
P_{N|\underline{s}_N} \left( \underline{y} | \underline{x} \right)^{\frac{1}{1+\rho}} \right]^{1 + \rho} \\
&= \sum_{\underline{y}} \left[ \sum_{\underline{x}}
\prod_{n=1}^N Q  (x_n) P \left( y_n | x_n, s_n \right)^{\frac{1}{1+\rho}} \right]^{1 + \rho} \\
&= \prod_{n=1}^N \sum_{y_n} \left[ \sum_{x_n}
Q  (x_n) P \left( y_n | x_n, s_n \right)^{\frac{1}{1+\rho}} \right]^{1 + \rho} ,
\end{split}
\end{equation*}
where $P_{N|\underline{s}_N} \left( \underline{y} | \underline{x} \right)=P (\underline{y}|\underline{x},\underline{s}_N)$ is the conditional distribution of receiving $\underline{y}$ given $\underline{x}$ and $\underline{s}_N$, and the first equality follows from (\ref{equation:ChannelDecomposition}). A key insight is to realize that this function only depends on $\underline{s}_N$ through its empirical distribution $T = \mathcal{T} \left( \underline{s}_N \right)$.
The proposition is then obtained by substituting this expression into equation (5.6.1) in \cite{Gallager-1968} and noticing that $e^{- \rho NR} = M^{\rho}$.
\end{IEEEproof}

\begin{corollary} \label{corollary:UpperBound}
Again, suppose Assumption~\ref{assumption:DecompositionRandomCodes} holds and let $u\subseteq\{\underline{s}_N'|\type(\underline{s}_N')=T\}$ such that $\Pr(u)>0$.
For any $\rho \in [0,1]$, the probability of decoding failure, conditioned on $\underline{s}_N\in u$ is also bounded by
\begin{equation*}
P_{\mathrm{e}|u}
\leq \exp \left( - N \left( E_{0,N} (\rho, \boldsymbol{Q}_N, \underline{s}_N) - \rho R \right) \right) .
\end{equation*}
\end{corollary}
\begin{IEEEproof}
Using the equivalence of the bound for all channel realizations with a same empirical distribution, we get
\begin{equation*}
\begin{split}
P_{\mathrm{e}|u} &= \sum_{\underline{s}_N \in u}
P_{\mathrm{e}|\underline{s}_N} \Pr \left( \underline{S}_N = \underline{s}_N| \underline{s}_N\in u \right) \\
&\leq \sum_{\underline{s}_N \in u}
e^{ - N \left( E_{0,N} (\rho, \boldsymbol{Q}_N, \underline{s}_N) - \rho R \right) }
\Pr \left( \underline{S}_N = \underline{s}_N| \underline{s}_N\in u \right) \\
&= e^{ - N \left( E_{0,N} (\rho, \boldsymbol{Q}_N, \underline{s}_N) - \rho R \right) } .
\end{split}
\end{equation*}
An interesting subset $u$ is one where the sequences start from a prescribed state $s_0$ and end in state $s_N$, while possessing the right empirical distribution.
\end{IEEEproof}


\begin{theorem}
Under Assumption~\ref{assumption:DecompositionRandomCodes}, the probability of decoding failure and ending in state $s_{N}$ conditioned on starting in
$s_{0}$, can be bounded as
\[
\sum_{T\in\type}\mbox{e}^{-N\left(E_{0,N}(\rho,\boldsymbol{\mbox{Q}}_{N},T)-\rho R\right)}P_{T,s_{N}|s_{0}},
\]
where $P_{T,s_{N}|s_{1}}$ denotes the probability of the empirical distribution of state sequence type ending in $s_{N}$ conditioned on the initial state $s_{1}$, and $E_{0,N}(\rho,\boldsymbol{\mbox{Q}}_{N},T)$ is given by Proposition~\ref{proposition:UpperBound}.
\end{theorem}
\begin{IEEEproof}
Following the same approach as in \cite[Section~5.9]{Gallager-1968},
we define
\begin{align*}
a(s_{n-1}\!,\!s_{n})\! & \triangleq\!\sum_{y_n}\!\{\!\sum_{x_n}\! Q(x_n)\!\left(\!P(\! s_{n}|s_{n-1}\!)P(\! y_n|x_n,s_{n-1}\!)\!\right)^{\frac{1}{1+\rho}}\!\}^{1+\rho}\\
 & =\!P(\! s_{n}|s_{n-1}\!)\!\sum_{j=0}^{J-1}\!\{\!\sum_{k=0}^{K-1}\! Q(k)P(\! j|k,s_{n-1}\!){}^{\frac{1}{1+\rho}}\!\}^{1+\rho}.
\end{align*}

By Proposition~\ref{proposition:UpperBound}, we have
\begin{align*}
\prod_{n=1}^{N}\!\!a(\!s_{n-1},s_{n}) & \!\!=\!\!\Pr \left( \underline{S}_N = \underline{s}_N|s_0\right)\mbox{exp}\!\left[-N\!\left(E_{0,N}\!\left(\rho,\boldsymbol{\mbox{Q}}_{N},\underline{s}_N\right)\!\right)\right]\!.
\end{align*}
Then we can write the upper bound on error probability as
\begin{align}
 & =\mbox{e}^{NR}\sum_{\underline{s}_N}\prod_{n=1}^{N}a(s_{n-1},s_{n})\nonumber \\
 & =\mbox{e}^{NR}\sum_{\underline{s}_N}\!\Pr \left( \underline{S}_N= \underline{s}_N|s_0\right)\mbox{e}^{-N\left(\!E_{0,N}(\rho,\boldsymbol{\mbox{Q}}_{N},\underline{s}_N)\!\right)}\nonumber \\
 & =\mbox{e}^{NR}\sum_{T\in\type}\sum_{\underline{s}_N\in T}\Pr \left( \underline{S}_N= \underline{s}_N|s_0\right)\mbox{e}^{-N\left(E_{0,N}(\rho,\boldsymbol{\mbox{Q}}_{N},\underline{s}_N)\right)}\nonumber \\
 & =\mbox{e}^{NR}\sum_{T\in\type}\mbox{e}^{-N\left(E_{0,N}(\rho,\boldsymbol{\mbox{Q}}_{N},\underline{s}_N)\right)}\sum_{\underline{s}_N\in T}\!\Pr \left( \underline{S}_N= \underline{s}_N|s_0\right)\nonumber \\
 & =\sum_{T\in\type}\mbox{e}^{-N\left(E_{0,N}(\rho,\boldsymbol{\mbox{Q}}_{N},\underline{s}_N)-\rho R\right)}P_{T,s_{N}|s_{0}}.
\label{eq:error bound}
\end{align}

\end{IEEEproof}

This upper bound consists of the exponential
upper bounds conditioned on the channel type, averaged over all state
sequence types. It means that for each type we have an exponential
decay with block length in the error probability. Note that $P_{\type,s_{N}|s_{0}}$
does not decay exponentially with $N$.

As an example, we now compute this upper bound for the Gilbert-Elliott
channel. Let $n_{\good}$ and $n_{\bad}=N-n_{\good}$ be the number of times that the
channel is in the good and bad states during the transmission
of a codeword of length $N$, respectively. These numbers are also
referred as the occupation times of the channel or the channel state
type (see chapter 12 in \cite{Cover-1991}). Then $\eta_{\good}\triangleq\frac{n_{\good}}{N}\,,\,\eta_{\bad}\triangleq\frac{n_{\bad}}{N}$ are the fraction of times in each state. The fractional type of the channel state sequence is another name to describe the fraction of times spent in each state. For example, type $(0.5,0.5)$ for a Gilbert-Elliott
channel means that the channel spent half of the times in each state .
By Proposition~\ref{proposition:UpperBound} we get
\begin{align}
E_{0,N}(\rho,\boldsymbol{\mbox{Q}}_{N},\underline{s}) & =-\left(\mbox{ln}G_{\bad}(\rho)\!+\!\eta_{\good}\mbox{ln}\frac{G_{\good}(\rho)}{G_{\bad}(\rho)}\right),
\label{eq:error exponent_gec}
\end{align}
where for the i.i.d input distribution,\vspace{-2mm}
\begin{align*}
G_{\good}(\rho) & =\frac{1}{2^{\rho}}\left(\epsilon_{\good}{}^{\frac{1}{1+\rho}}+(1-\epsilon_{\good})^{\frac{1}{1+\rho}}\right)^{1+\rho}\\
G_{\bad}(\rho) & =\frac{1}{2^{\rho}}\left(\epsilon_{\bad}{}^{\frac{1}{1+\rho}}+(1-\epsilon_{\bad})^{\frac{1}{1+\rho}}\right)^{1+\rho}
\end{align*}

Also by small modification to Gallager's derivation for $E_{0,N}(\rho,\boldsymbol{\mbox{Q}}_{N},s_{0})$, ((5.9.39) in \cite{Gallager-1968}), we have
\begin{align*}
P_{e,s_N|s_0}&<\!\min_{0\leq\rho\leq1}\!\left\{\!\left(e(s_1)\!\left[\!\begin{array}{cc}
a(\good,\good) & a(\good,\bad)\\
a(\bad,\good) & a(\bad,\bad)
\end{array}\right]^{N}\!e(s_N)\!\right)\!\!\mbox{e}^{\rho NR}\!\!\right\}\!\!,
\end{align*}
where $e(s_i)$ is the unit vector with a one in the $i$-th position, and $a(\good,\good)=(1-\alpha) G_{\good}(\rho)$, $a(\good,\bad)=\alpha G_{\good}(\rho)$, $a(\bad,\good)=\beta G_{\bad}(\rho)$, and $a(\bad,\bad)=(1-\beta)G_{\bad}(\rho)$. Notice that this bound is the same as the bound computed by the sum given in (\ref{eq:error bound}) using (\ref{eq:error exponent_gec}), and the joint distribution of $\eta_{\good}$ and $s_{N}$, conditioned on $s_{0}$.

On the other hand in the rare-transition regime in which $N\alpha$ and $N\beta$ are constant, by taking the limit as $N\rightarrow\infty$, we can compute the bound on failure probability (\ref{eq:error bound}), with a small error (see \cite{Resnick-99}), as follows
\begin{align}
& \min_{0\leq\rho\leq1}\int_{0}^{1}\!\!\!\mbox{e}^{N\!\left(\!\mbox{ln}G_{\bad}(\rho)+x\mbox{ln}\frac{G_{\good}(\rho)}{G_{\bad}(\rho)}+\rho R\!\right)}f(x,s_{N}=d|s_{0}=c)dx\nonumber \\
& =\min_{0\leq\rho\leq1}\left[\mbox{e}^{N\!\left(\!\mbox{ln}G_{\bad}(\rho)+\rho R\right)}\mathbf{G}_{c d}\left(N\mbox{ln}\frac{G_{\good}(\rho)}{G_{\bad}(\rho)}\right)\right],\label{eq:int}
\end{align}
where, $\mathbf{G}_{cd}(.)$, $c,d\in\{\good,\bad\}$ is the moment generating function of the limiting occupancy time distribution and is given by (\ref{eq:gf}).
By modifying Gallager's approach and considering the rare-transition regime, we get an upper bound on error probability of maximum-likelihood decoding that retains its dependency on the initial and final states.

\vspace{-3mm}
\section{Derivation of Exact Decoding Error Probability}
In this section, we first present the exact expression for failure probability over BSC and then derive failure probability over Gilbert-Elliott channel to compare with the upper bound.
\vspace{-5mm}
\subsection{Random Coding Error Probability for the BSC}

In \cite{Fano-1961}, the error probability for random coding over BSC($p$) using maximum-likelihood (ML) decoder, for a system which treats ties as error is derived as
\vspace{-1.5mm}
\begin{align*}
P_{e} & =\!\sum_{\tau=0}^{N}\!\binom{N}{\tau}p^{\tau}\!(\!1\!-\! p\!)^{N-\tau}\!\!\!\left[\!1\!\!-\!\!\left(\!\!1\!-\!2^{-N}\!\sum_{j=0}^{\tau}\!\binom{N}{j}\!\!\!\right){}^{\! M-1}\!\right]\\
& =\!1\!-2^{N\!-\! NM}\!\sum_{\tau=0}^{N}\!\binom{N}{\tau}p^{\tau}\!(\!1\!-\! p\!)^{N-\tau}\!\!\!\left(\!\sum_{j=\tau+1}^{N}\!\binom{N}{j}\!\!\!\right)^{\! M-1}
 \end{align*}
In \cite{Polyanskiy-it10}, some small modifications has been made to the Fano's expression to take ties into account as\vspace{-1mm}
\begin{align*}
P_{e}=1-2^{N-NM}\sum_{\tau=0}^{N}\binom{N}{\tau}p^{\tau}(1-p)^{N-\tau}\,\,\,\,\,\,\,\,\,\,\,\,\,\,\,\,\,\,\,\,\,\,\,\,\,\,\,\,\,\,\,\,\,\,\,\,\,\,\,\,\,\,\,\,\,\,\,\\
\times\left(\sum_{l=0}^{M-1}\binom{N}{\tau}^{l}\frac{1}{l+1}\binom{M-1}{l}\left(\sum_{j=\tau+1}^{N}\binom{N}{j}\right)^{M-1-l}\right).\label{eq:verdu}
\end{align*}

One can see that this gives a small difference in the error probability conditioned on the number of errors.\vspace{-3mm}
\subsection{The Gilbert-Elliott Channel: State Known at the Receiver}

Now, we consider data transmission over the Gilbert-Elliott channel using random coding when the state is known at the receiver. Two different decoders are considered: a minimum distance decoder and a maximum-likelihood decoder. There are some differences between these two decoding rules which we will go
through in detail in the following subsections. We then derive the
decoding failure probability for the two decoders, conditioned
on the knowledge of the occupancy times. It turns out that
when the state is known at the receiver, the empirical distribution of the channel state provides enough information to determine the error probability.
Using the distribution of $n_{\good}$ and the $n_{\good}$-conditional error probabilities for different decoding rules, one can average over all types to get the probability of decoding failure,\vspace{-1.5mm}
\begin{equation}
P_{e}=\sum_{T\in\type}P_{e|T}\Pr(\underline{s}_N\in T),\label{eq:1}
\end{equation}
where $P_{e|T}$ is the probability of decoding error given that the channel state sequence type is $T$, and $\Pr(\underline{s}_N\in T)$ denotes the probability distribution of channel type $T$ which will be derived in the following section.
\subsubsection{Minimum Distance Decoding}
Given the channel type, the minimum distance decoder on
the Gilbert-Elliott channel acts similar to a maximum likelihood decoder on BSC.

Suppose we denote the number of errors in each
state by $e_{\good}$ and $e_{\bad}$, where $e_{\good}=d_{H}\left(\underline{X}_{\good},\underline{Y}_{\good}\right)$ and $e_{\bad}=d_{H}\left(\underline{X}_{\bad},\underline{Y}_{\bad}\right)$.
The conditional error probability can be written as
\begin{equation}
\begin{array}{cc}
P_{e|T} & =\sum_{e_{\good}=0}^{n_{\good}}\sum_{e_{\bad}=0}^{n_{\bad}}P_{e|T,e_{\good},e_{\bad}}P_{e_{\good},e_{\bad}|T},\end{array}\label{eq:2}
\end{equation}
where $P_{e|T,e_{\good},e_{\bad}}$ is the error probability conditioned on type $T$ and the number of errors in each state, and $P_{e_{\good},e_{\bad}|T}$ is the probability oh having $e_\good$ and $e_\bad$ errors conditioned on $T$. Conditioned on the channel type, the number of errors in the good
and bad states are independent. So,
\begin{equation}
P_{e_{\good},e_{\bad}|\type}=P_{e_{\good}|\type}P_{e_{\bad}|\type},\label{eq:3-1}
\end{equation}
and we have\vspace{-1mm}
\begin{equation}
\begin{array}{cc}
&P_{e_{\good}|\type}=\!\binom{n_{\good}}{e_{\good}}\epsilon_{\good}^{e_{\good}}(1-\epsilon_{\good})^{n_{\good}-e_{\good}}, \\ &P_{e_{\bad}|\type}=\!\binom{n_{\bad}}{e_{\bad}}\epsilon_{\bad}^{e_{\bad}}(1-\epsilon_{\bad})^{n_{\bad}-e_{\bad}},
\end{array}
\label{eq:3}
\end{equation}
\vspace{-2mm}
\begin{equation}
P_{e|\type,e_{\good},e_{\bad}}=P_{e|e_{\good},e_{\bad}}.\label{eq:4}
\end{equation}
\vspace{-.5mm}
Because conditioned on $e_{\good}$ and $e_{\bad}$, the probability of error is
independent of the channel state sequence type.

If there exists at least one codeword other than
the transmitted one inside the decoding sphere of radius $e_{\good}+e_{\bad}$
centered at the received word, the bounded distance decoder either decodes
to an incorrect codeword or fails to decode. This means ties
and errors are grouped. Since there are $M-1$ codewords other than the transmitted
one, the probability that none of them falls inside the decoding sphere
is $\left(1-2^{-N}\sum_{j=0}^{e_{\good}+e_{\bad}}\binom{N}{j}\right)^{M-1}$
and the probability that at least one of them falls inside the ball
is simply
\begin{equation}
P_{e|\type,e_{\good},e_{\bad}}=1-\left(1-2^{-N}\sum_{j=0}^{e_{\good}+e_{\bad}}\binom{N}{j}\right)^{M-1}.\label{eq:6}
\end{equation}

Now, we can simply substitute all the terms in equation \ref{eq:2}
to have the error probability expression.

To take into account the
ties, and not to treat all the tie events as error, we can slightly
modify the equation \ref{eq:6} in the same way as \ref{eq:verdu}.

\subsubsection{\label{sub:ML}Maximum-Likelihood Decoding}
\begin{lemma}
When the state is known at the receiver, the maximum-likelihood decoder decodes to the following codeword\vspace{-2mm}
\begin{align*}
\underset{\underline{X}\in\mathcal{C}}{\arg \max}\,\,\ln(P(\underline{Y}|\underline{X})) & =\underset{\underline{X}\in\mathcal{C}}{\arg \min}[\left\lceil\gamma e_{\good}\right\rceil+e_{\bad}],
\end{align*}
where~$\gamma=\frac{\ln\epsilon_{\good}-\ln(1-\epsilon_{\good})}{\ln\epsilon_{\bad}-\ln(1-\epsilon_{\bad})}>1$. Moreover, the error probability conditioned on the number of errors in each state and the channel state type is\vspace{-2mm}
\begin{equation}
1-\left(1-2^{-N}\sum_{\left\lceil\gamma\tilde{e}_{\good}\right\rceil+\tilde{e}_{\bad}\leq C}\binom{n_{\good}}{\tilde{e}_{\good}}\binom{N-n_{\good}}{\tilde{e}_{\bad}}\right)^{M-1},\label{eq:7}
\end{equation}
where $C\triangleq\left\lceil\gamma e_{\good}\right\rceil+e_{\bad}$
\label{lem:ML}
\end{lemma}
\begin{IEEEproof}
The proof is given in \ref{sub:mlProof}.
\end{IEEEproof}

\vspace{-3.5mm}
\section{Distribution of Channel State Type for Gilbert-Elliott Channel}

The purpose of the current analysis is to study delay-sensitive communication
systems and evaluate the queueing behavior. Since in these
kind of systems the block-length cannot go extremely large, it turns
out that the effect of the initial and the final states does not go
away with $N$ for moderate block lengths. So the error probability
depends on the initial and the final states ($s_{0}$ and $s_{N}$,
respectively). Conditioned on the channel type, the error probability
is independent of the initial and the final states. The only part
that depends on these states is the distribution of the state occupancies.
In \cite{Gabriel-Biometrika59,Pedler-jap71}, the probability distribution of the occupation
times for two-state Markov chains is derived. However, the given distribution is averaged over all final states. Doing some manipulation we will derive the joint probability
distribution of state occupation $n_{\good}$ and final state conditioned
on the initial state.

\begin{theorem}
The joint distribution of channel type and the final state conditioned on the initial final state can be computed as\vspace{-1mm}
\begin{align*}
\mbox{Pr}(n_{\good}=m,s_{N}=\good|s_{0}=\good) & =(1-\alpha)^{m}(1-\beta)^{N-m}\times\\
\left\{ F(-N\!+\! m,-m;1;\lambda)\!\right. & \!\left.-F(-N\!+\! m\!+\!1,-m;1;\lambda)\right\},\\
\mbox{Pr}(n_{\good}=m,s_{N}=\bad|s_{0}=\good) & =\frac{(1-\alpha)^{m-1}(1-\beta)^{N-m+1}\alpha}{(1-\beta)}\\
 & \,\,\times F(-N+m,-m+1;1;\lambda),\\
\mbox{Pr}(n_{\good}=m,s_{N}=\good|s_{0}=\bad) & =\frac{(1-\alpha)^{m+1}(1-\beta)^{N-m-1}\beta}{(1-\alpha)}\\
 & \,\,\times F(-N+m+1,-m;1;\lambda),\\
\mbox{Pr}(n_{\good}=m,s_{N}=\bad|s_{0}=\bad) & =(1-\alpha)^{m}(1-\beta)^{N-m}\times\\
\left\{ F(-N\!+\! m,-m;1;\lambda)\!\right. & \!\left.-F(-N\!+\! m,-m\!+\!1;1;\lambda)\right\},
\end{align*}
for $0<m<N$. $F(.,.;.)$ is the hypergeometric function, $\lambda=\frac{\mathbf{\alpha\beta}}{(1-\alpha)(1-\beta)}$,
and $d=1-\alpha-\beta$. Moreover, $\mbox{Pr}(n_{\good}=0,s_{N}|s_{0}=\good)=0$, $\mbox{Pr}(n_{\good}=T,s_{N}|s_{0}=\bad)=0$, $\mbox{Pr}(n_{\good}=0,s_{N}=\bad|s_{0}=\bad)=(1-\beta)^{N}$, and $\mbox{Pr}(n_{\good}=N,s_{N}=\good|s_{0}=\good)=(1-\alpha)^{N}$.
\label{thm:1}
\end{theorem}
\begin{IEEEproof}
Proof of the theorem is given in \ref{sub:dtmc}.
\end{IEEEproof}
Notice that we can also compute these probabilities using the generating matrix method. Let\vspace{-2.1mm}
\[
\mathbf{G}_{\good}(x)=\left[\begin{array}{cc}
(1-\alpha)x & \alpha x\\
\beta  & 1-\beta
\end{array}\right].
\]
By taking the $N$-th power of this matrix, coefficient of the $m$-th power of $x$ in the corresponding entry represents the above conditional probability. However, this method does not give us the closed form distribution of the occupation times, directly.

To get the distribution of the fractional occupation time $\eta_{\good}$, we consider the rare-transition regime in which the transition probabilities are scaled with $N$ as $\alpha_{N}=\frac{\alpha}{N}$ and $\beta_{N}=\frac{\beta}{N}$. In this regime, the expected number of transitions in each length-$N$ block is constant. Taking the limit of the above conditional probabilities for the DTMC as $N\rightarrow\infty$ gives us the distribution of the fractional occupancy time $x=\lim_{N\rightarrow\infty}\frac{m}{N}$.
Notice that the transition-rate matrix of the CTMC is $\mathbf{Q}=\left[\begin{array}{cc}
-\alpha & \alpha\\
\beta & -\beta
\end{array}\right]$.
For the DTMC in the rare transitions regime, the probability transition matrix is\vspace{-1.8mm}
\[
\mathbf{P}_{N}=\left[\begin{array}{cc}
1-\frac{\alpha}{N} & \frac{\alpha}{N}\\
\frac{\beta}{N} & 1-\frac{\beta}{N}
\end{array}\right]=\mathbf{I}+\frac{1}{N}\mathbf{Q},
\]
where $\mathbf{I}$ is the 2 by 2 identity matrix. So,
\begin{align*}
\lim_{N\rightarrow\infty}\left(\mathbf{P}_{N}\right)^{Nt} & =\left(\mathbf{I}+\frac{1}{N}\mathbf{Q}\right)^{Nt}=\mbox{e}^{\mathbf{Q}t}.
\end{align*}
This means that the rare-transition limit of the DTMC results yields the corresponding results for the CTMC.
Similarly,\vspace{-1.5mm}
\begin{equation}
\left(\begin{array}{cc}
\mathbf{G}_{\good\good}(y) & \mathbf{G}_{\good\bad}(y)\\
\mathbf{G}_{\bad\good}(y) & \mathbf{G}_{\bad\bad}(y)
\end{array}\right)\triangleq\lim_{N\rightarrow\infty}\mathbf{G}_{\good,N}\left(1+\frac{y}{N}\right),\label{eq:gf}
\end{equation}
where $\mathbf{G}_{\good,N}$ is obtained by replacing $\alpha$ and $\beta$ in $\mathbf{G}_{\good}(.)$ with $\alpha_{N}$ and $\beta_{N}$, gives the corresponding matrix generating function.
\vspace{-2mm}
\begin{lemma}
The joint distribution of fractional occupancy times and the final state conditioned on the initial state can be computed as\vspace{-0.6mm}
\begin{align*}
f(x,s_{N}=\good|s_{0}=\good) & =e^{-\alpha x-\beta(1-x)}\{\delta(1-x) \\
 & \,\,\,+\left(\frac{\alpha\beta x}{1-x}\right)^{\frac{1}{2}}I_{1}[2(\alpha\beta x(1-x))^{\frac{1}{2}}]\},\\
f(x,s_{N}=\bad|s_{0}=\good) & =\alpha e^{-\alpha x-\beta(1-x)}I_{0}[2(\alpha\beta x(1-x))^{\frac{1}{2}}],\\
f(x,s_{N}=\good|s_{0}=\bad) & =\beta e^{-\alpha x-\beta(1-x)}I_{0}[2(\alpha\beta x(1-x))^{\frac{1}{2}}],\\
f(x,s_{N}=\bad|s_{0}=\bad) & =e^{-\alpha x-\beta(1-x)}\{\delta(x) \\
 & \,\,\,+\left(\frac{\alpha\beta (1-x)}{x}\right)^{\frac{1}{2}}I_{1}[2(\alpha\beta x(1-x))^{\frac{1}{2}}]\}.
\end{align*}
\label{lem:1}\vspace{-3mm}
\end{lemma}
\begin{IEEEproof}
The proof can be found in \ref{sub:ctmc}.
\end{IEEEproof}


\vspace{-3mm}
\section{Numerical Results}
\begin{figure}
\begin{center}
%
%
\begin{tikzpicture}[scale=0.8]

\begin{semilogyaxis}[%
scale only axis,
width=3in,
height=1.5in,
xmin=0.25, xmax=0.8,
ymin=1e-005, ymax=1,
yminorticks=true,
xmajorgrids,ymajorgrids,
xlabel={Rate},
ylabel={Failure Probability},
axis on top,
legend entries={\small{Gallager's Bound} $N=50$,\small{Gallager's Bound} $N=75$,\small{Gallager's Bound} $N=100$,\small{Rare - transition} $N=50$,\small{Rare - transition} $N=75$,\small{Rare - transition} $N=100$},
legend style={at={(1,0)},anchor={south east}}]
\addplot [
color=black,
dashed,
line width=1.0pt,
mark=o,
mark options={solid}
]
coordinates{
 (0.25,0.000624627)(0.3,0.00290582)(0.35,0.0105345)(0.4,0.0310782)(0.45,0.076987)(0.5,0.163529)(0.55,0.302026)(0.6,0.489344)(0.65,0.698883)(0.7,0.881017)(0.75,0.981398) 
};

\addplot [
color=black,
dashed,
line width=1.0pt,
mark=asterisk,
mark options={solid}
]
coordinates{
 (0.25,8.13177e-005)(0.3,0.000614999)(0.35,0.00324139)(0.4,0.0128583)(0.45,0.0403769)(0.5,0.103789)(0.55,0.223285)(0.6,0.407658)(0.65,0.636352)(0.7,0.850843)(0.75,0.975434) 
};

\addplot [
color=black,
dashed,
line width=1.0pt,
mark=triangle,
mark options={solid,,rotate=180}
]
coordinates{
 (0.25,1.35385e-005)(0.3,0.000163962)(0.35,0.00122813)(0.4,0.00635319)(0.45,0.0244339)(0.5,0.0734126)(0.55,0.178083)(0.6,0.35614)(0.65,0.594018)(0.7,0.829353)(0.75,0.971017) 
};

\addplot [
color=black,
solid,
line width=1.0pt,
mark=o,
mark options={solid}
]
coordinates{
 (0.25,0.000792121)(0.3,0.00353258)(0.35,0.0122949)(0.4,0.0349733)(0.45,0.0839249)(0.5,0.173525)(0.55,0.313498)(0.6,0.499286)(0.65,0.704336)(0.7,0.881136)(0.75,0.979541) 
};

\addplot [
color=black,
solid,
line width=1.0pt,
mark=asterisk,
mark options={solid}
]
coordinates{
 (0.25,0.000100155)(0.3,0.000732905)(0.35,0.00373933)(0.4,0.0143839)(0.45,0.0439147)(0.5,0.11012)(0.55,0.231987)(0.6,0.416473)(0.65,0.642044)(0.7,0.851627)(0.75,0.973924) 
};

\addplot [
color=black,
solid,
line width=1.0pt,
mark=triangle,
mark options={solid,,rotate=180}
]
coordinates{
 (0.25,1.62288e-005)(0.3,0.000191721)(0.35,0.00139895)(0.4,0.00705018)(0.45,0.0264491)(0.5,0.0776962)(0.55,0.184833)(0.6,0.363783)(0.65,0.599496)(0.7,0.830454)(0.75,0.969766) 
};

\end{semilogyaxis}

\end{tikzpicture}
\end{center}
\vspace{-5mm}
\caption{Comparison of our derived bound in (\ref{eq:int}) with Gallager-type bound (\ref{eq:error bound}) in rare-transition regime for $N\alpha=0.04$ and $N\beta=0.12$}
\label{figure:bounds}
\end{figure}
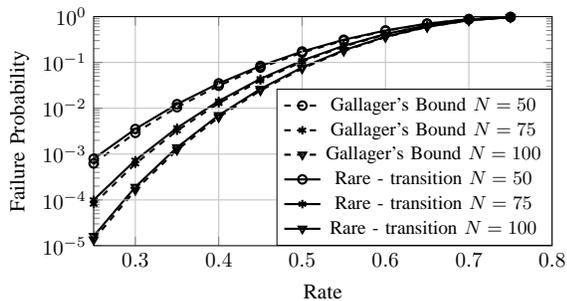

\begin{figure}
\begin{center}
%
%
\begin{tikzpicture}[scale=0.8]

\begin{semilogyaxis}[%
scale only axis,
width=3in,
height=1.5in,
xmin=0.25, xmax=0.8,
ymin=1e-006, ymax=1,
yminorticks=true,
xmajorgrids,ymajorgrids,
xlabel={Rate},
ylabel={Failure Probability},
axis on top,
legend entries={$\text{ML N}=50$,$\text{ML N}=75$,$\text{MD N}=50$,$\text{MD N}=75$,$\text{Bound N}=50$,$\text{Bound N}=75$},
legend style={at={(1,0)},anchor={south east}}]
\addplot [
color=black,
solid,
line width=1.0pt,
mark=o,
mark options={solid}
]
coordinates{
 (0.25,0.000178344)(0.3,0.00067079)(0.35,0.00214493)(0.4,0.00591513)(0.45,0.0143075)(0.5,0.0308196)(0.55,0.0601084)(0.6,0.107283)(0.65,0.175714)(0.7,0.267464)(0.75,0.383104) 
};

\addplot [
color=black,
solid,
line width=1.0pt
]
coordinates{
 (0.25,6.89775e-006)(0.3,4.99e-005)(0.35,0.000276952)(0.4,0.00120856)(0.45,0.00428744)(0.5,0.0127238)(0.55,0.0323583)(0.6,0.0716219)(0.65,0.139908)(0.7,0.243539)(0.75,0.380866) 
};

\addplot [
color=black,
dashed,
line width=1.0pt,
mark=o,
mark options={solid}
]
coordinates{
 (0.25,0.000314153)(0.3,0.00115953)(0.35,0.00365273)(0.4,0.00998224)(0.45,0.0235163)(0.5,0.0480992)(0.55,0.087804)(0.6,0.149932)(0.65,0.2495)(0.7,0.344088)(0.75,0.445695) 
};

\addplot [
color=black,
dashed,
line width=1.0pt
]
coordinates{
 (0.25,1.53958e-005)(0.3,0.000110869)(0.35,0.000609978)(0.4,0.00273185)(0.45,0.00855393)(0.5,0.0250595)(0.55,0.0599197)(0.6,0.115746)(0.65,0.203125)(0.7,0.322932)(0.75,0.469847) 
};

\addplot [
color=black,
dash pattern=on 1pt off 3pt on 3pt off 3pt,
line width=2.0pt,
mark=o,
mark options={solid}
]
coordinates{
 (0.25,0.0018571)(0.3,0.0071387)(0.35,0.0215592)(0.4,0.0538627)(0.45,0.11517)(0.5,0.215584)(0.55,0.358613)(0.6,0.535127)(0.65,0.71997)(0.7,0.874625)(0.75,0.967494) 
};

\addplot [
color=black,
dash pattern=on 1pt off 3pt on 3pt off 3pt,
line width=2.0pt
]
coordinates{
 (0.25,0.000100155)(0.3,0.000732905)(0.35,0.00373933)(0.4,0.0143839)(0.45,0.0439147)(0.5,0.11012)(0.55,0.231987)(0.6,0.416473)(0.65,0.642044)(0.7,0.851627)(0.75,0.973924) 
};

\end{semilogyaxis}

\end{tikzpicture}
\end{center}\vspace{-5mm}
\caption{Comparison of our derived bound in (\ref{eq:int}) with the exact values for maximum-likelihood and minimum distance decoders for $\alpha=0.0533$ and $\beta=0.08$}
\label{figure:exactVSbounds}
\end{figure}
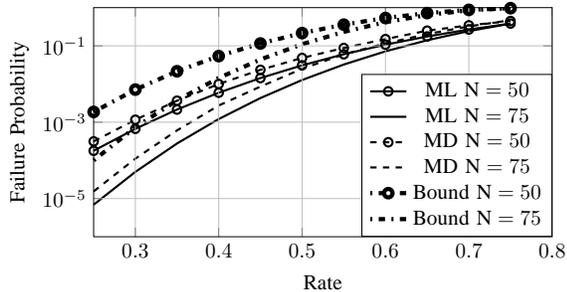

In this section we present the numerical results for a system which transmits data over the Gilbert-Elliott channel with  $\epsilon_{\good}=0.01$, $\epsilon_{\bad}=0.1$. Fig. \ref{figure:bounds} represents our derived bound in (\ref{eq:int}) with Gallager-type bound (\ref{eq:error bound}) in rare-transition regime where $N\alpha=0.04$, and $N\beta=0.06$. The plots show the averaged failure probability over all state transitions. As we can see, although the block-lengths are short, the bounds given by (\ref{eq:int}) are very close to (\ref{eq:error bound}) while keeping a simpler format. In Fig. \ref{figure:exactVSbounds} we compare the exact results for maximum-likelihood and minimum distance decoders with our derived upper bound given by (\ref{eq:int}) for fixed transitions probabilities. As we expect, maximum-likelihood decoder outperforms minimum distance decoder. Moreover, by increasing the block-length the bound gets closer to the exact value.
\vspace{-3mm}

\vspace{-1mm}

\section{Conclusion}
We proposed a general approach to bound the failure probability for random coding over finite-state channels which retains the dependency on the initial and final states for relatively short block-lengths. We also derived expressions to compute the exact error probability for maximum-likelihood and minimum distance decoders for Gilbert-Elliott channel and we compared them with the proposed upper bound.

\vspace{-3mm}
\bibliographystyle{ieeetr}
\bibliography{IEEEabrv,IEEEexample,IEEEfull,WCLnewbib,WCLfull,WCLbib,WCLabrv}

\section{Appendix}
\subsection{Proof of Lemma~\ref{lem:ML}}\label{sub:mlProof}
First we revisit the ML decoding rule for the Gilbert-Elliott channel,
when the state is known at the receiver and conditioned on the channel
state type ($n_{\good}$). So, we have
\[
P(\underline{Y}|\underline{X})=\epsilon_{\good}^{e_{\good}}(1-\epsilon_{\good})^{n_{\good}-e_{\good}}\epsilon_{\bad}^{e_{\bad}}(1-\epsilon_{\bad})^{N-n_{\good}-e_{\bad}}
\]
Upon receiving the word $\underline{Y}$, the ML decoder decodes to the codeword
$\underline{X}$ that maximizes $P(\underline{Y}|\underline{X})$. Equivalently,
the decoded message will be $\underset{\underline{X}\in\mathcal{C}}{\mbox{arg min}}[\gamma e_{\good}+e_{\bad}]$.
It means that for the ML decoder, the errors in the bad state
do not cost the system as much as the errors in the good state. That
is because the receiver expects errors more in
the bad state.

To get the error probability for ML decoder, similar to the minimum distance decoder, we first condition on $e_{\good}$
and $e_{\bad}$. Then for each set of $e_{\good}$ and $e_{\bad}$ we define $C\triangleq\left\lceil\gamma e_{\good}\right\rceil+e_{\bad}$. Then, the probability
of error conditioned on $n_{\good},\, e_{\good},\,\mbox{and}\, e_{\bad}$ is obtained as (\ref{eq:7}).
\subsection{Proof of Theorem~\ref{thm:1}}\label{sub:dtmc}
From \cite{Pedler-jap71} we have
\begin{align*}
\mbox{Pr}(n_{\good}\!=\! m|s_{0}\!=\!\good) & =(1\!-\!\alpha)^{m}(1\!-\!\beta)^{N\!-m}\!\sum_{c=0}^{c_{1}}\!\binom{m}{a}\!\binom{\!N\!-\!m\!-\!1}{b\!-\!1}\\
 & \times\left(\frac{\alpha}{1-\beta}\right)^{b}\left(\frac{\beta}{1-\alpha}\right)^{a},
\end{align*}

where $a$ and $b$ are the number of transitions into the initial
state and out of the initial state, respectively. $c$ is the total
number of transitions which occur up to time $N$, and
\[
c_{1}=\begin{cases}
N+\frac{1}{2}-\left|2m-\frac{1}{2}+N\right|, & m<N\\
0, & m=N
\end{cases}.
\]
By splitting the summation into terms for which $c$ is odd and $c$
is even, it follows that if $c=2k$, $a=b=k$
and the corresponding sum represents $\mbox{Pr}(n_{g}\!=\! m,s_{N}\!=\good|s_{0}\!=\good)$.
If $c=2k+1$, $a=k$, $b=k+1$, and the corresponding sum represents
$\mbox{Pr}(n_{g}\!=\! m,s_{N}\!=\bad|s_{0}\!=\good)$. So we have
\begin{align*}
 & \mbox{Pr}(n_{\good}\!=\! m|s_{0}\!=\!\good)=\\
 & (\!1\!-\!\alpha\!)^{m}(\!1\!-\!\beta)^{N\!- m}\!\sum_{k}\!\binom{m}{k}\!\!\binom{N\!-\! m\!-\!1}{k\!-\!1}\!\!\!\left(\!\!\frac{\alpha}{1\!-\!\beta}\!\!\right)^{\!\! k}\!\!\left(\!\!\frac{\beta}{1\!-\!\alpha}\!\!\right)^{\!\! k}\\
 & +\!(\!1\!-\!\alpha\!)^{m}(\!1\!-\!\beta)^{N\!- m}\!\sum_{k}\!\binom{m}{k}\!\!\binom{N\!-\! m\!-\!1}{k}\!\!\!\left(\!\!\frac{\alpha}{1\!-\!\beta}\!\!\right)^{\!\! k+1}\!\!\!\left(\!\!\frac{\beta}{1\!-\!\alpha}\!\!\right)^{\!\! k}
\end{align*}

We can set the upper and lower limit on $k$ to $0$ and $\infty$,
since all other terms are automatically zero. From the definition of $F(-N\!+\! m\!+\!1,-m;1;\lambda)$ we see that
\begin{align*}
  \!\left(\!\!\frac{\alpha}{1\!-\!\beta}\!\!\right)\!\sum_{k=0}^{\infty}\!\!\binom{m}{k}\!\!\binom{N\!-\! m\!-\!1}{k}\!\!\!\left(\!\!\frac{\alpha}{1\!-\!\beta}\frac{\beta}{1\!-\!\alpha}\!\!\right)^{\!\!k}&\\
  =\!\left(\!\!\frac{\alpha}{1\!-\!\beta}\!\!\right)\!F(-N\!+\! m\!+\!1,-m;1;\lambda)
\end{align*}
So we have
\begin{align*}
\mbox{Pr}(n_{\good}\!=\! m,s_{N}\!=\bad|s_{0}\!=\good) & =(\!1\!-\!\alpha\!)^{m}(\!1\!-\!\beta\!)^{N\!-m}\times\\
& \!\left(\!\!\frac{\alpha}{1\!-\!\beta}\!\!\right)\!F(-N\!+\! m\!+\!1,-m;1;\lambda),
\end{align*}
for $0<m<N$. Clearly, for $m=0$ and $m=N$, this conditional probability equals to $0$.

By \cite{Pedler-jap71} and noticing that
\begin{align*}
\mbox{Pr}(n_{\good}\!=\! m|s_{0}\!=\good) & =\mbox{Pr}(n_{\good}\!=\! m,s_{N}\!=\good|s_{0}\!=\good)\\
& +\mbox{Pr}(n_{\good}\!=\! m,s_{N}\!=\bad|s_{0}\!=\good),
\end{align*}
we have
\begin{align*}
\mbox{Pr}(n_{\good}\!=\! m,s_{N}\!=\good|s_{0}\!=\good) & =(\!1\!-\!\alpha\!)^{m}(\!1\!-\!\beta)^{N\!-m}\times\\
& \left\{ \!F(-N\!+\! m,-m;1;\lambda)\!-\!d(1\!-\!\beta)\!^{-1}\right.\\
& \left.\times F(-N\!+\! m\!+\!1,-m;1;\lambda)\!\right\}\\
& -(\!1\!-\!\alpha\!)^{m}(\!1\!-\!\beta)^{N\!-m}\times\\
& \!\left(\!\!\frac{\alpha}{1\!-\!\beta}\!\!\right)\!F(-N\!+\! m\!+\!1,-m;1;\lambda)\\
& =(\!1\!-\!\alpha\!)^{m}(\!1\!-\!\beta)^{N\!-m}\times\\
& \left\{ F(-N\!+\! m,-m;1;\lambda)\right.\\
& \left.-F(-N\!+\! m\!+\!1,-m;1;\lambda)\right\},
\end{align*}
for $0<m<N$. Moreover, $\mbox{Pr}(n_{\good}\!=\!0,s_{N}\!=\good|s_{0}\!=\good)=0$ and $\mbox{Pr}(n_{\good}\!=\!N,s_{N}\!=\good|s_{0}\!=\good)=(1-\alpha)^{N}$.

The other conditional probabilities can be derived in the same manner.
\subsection{Proof of Lemma~\ref{lem:1}}\label{sub:ctmc}
In \cite{Pedler-jap71}, Pedler considers
a Markov chain with two states and continuous time parameter set $[0,t].$
Then he defines the occupation time $X(t)$ as the time spent in the
first state (the good state) during the interval $[0,t]$, and derives
the PDF (probability density function) of $X(t)$, called $f(x,t)$.
It has been shown that
\begin{align*}
f(x,t) & =e^{-\alpha x-\beta(t-x)}\{\pi_{g}\delta(t-x)+\pi_{b}\delta(x)\\
 & +[\pi_{g}(\frac{\alpha\beta x}{t-x})^{\frac{1}{2}}+\pi_{b}(\frac{\alpha\beta(t-x)}{x})^{\frac{1}{2}}]I_{1}[2(\alpha\beta x(t-x))^{\frac{1}{2}}]\\
 & +(\pi_{g}\alpha+\pi_{b}\beta)I_{0}[2(\alpha\beta x(t-x))^{\frac{1}{2}}],
\end{align*}
where $\pi_{\good}$ and $\pi_{\bad}$ are the steady state probability of
being in the good state and the bad state, respectively. $I_{0}(.)$
and $I_{1}(.)$ are the modified Bessel functions of order 0 and 1,
respectively.

First, we put $t=1$ in this formula to normalize the time interval
and have the distribution of fractional occupancy times with respect
to $N$, $f(x)$. Then we rewrite this PDF as
\begin{align*}
f(x) & =\pi_{\good}f(x|s_{0}=\good)+\pi_{\bad}f(x|s_{0}=\bad)\\
 & =\pi_{\good}[f(x,s_{N}=\good|s_{0}=\good)\\
 & \,\,\,\,\,\,\,\,+f(x,s_{N}=\bad|s_{0}=\good)]\\
 & \,\,\,\,\,\,\,\,+\pi_{\bad}[f(x,s_{N}=\good|s_{0}=\bad)\\
 & \,\,\,\,\,\,\,\,+f(x,s_{N}=\bad|s_{0}=\bad)].
\end{align*}

Now, we want to compute
\begin{align*}
& \lim_{N\rightarrow\infty}\,\mbox{Pr}(n_{\good}\!=\! m,s_{N}\!=\bad|s_{0}\!=\good)=\\
& \lim_{N\rightarrow\infty}\!\left[\!(\!1\!-\!\frac{\alpha}{N}\!)^{m}(\!1\!-\!\frac{\beta}{N}\!)^{N\!-m}\!\left(\!\!\frac{\frac{\alpha}{N}}{1\!-\!\frac{\beta}{N}}\!\!\right)\!F(-N\!+\! m\!+\!1,-m;1;\lambda^{\prime})\!\right]\!.
\end{align*}
where $\lambda_{N}=\frac{\frac{\alpha}{N}\frac{\beta}{N}}{\left(1-\frac{\alpha}{N}\right)\left(1-\frac{\beta}{N}\right)}$. First consider
\begin{align*}
& \lim_{N\rightarrow\infty}F\left(-N\!+\! m\!+\!1,-m;1;\frac{\frac{\alpha}{N}\frac{\beta}{N}}{\left(1\!-\!\frac{\alpha}{N}\right)\left(1\!-\!\frac{\beta}{N}\right)}\right).
\end{align*}
By the definition of $F(-N\!+\! m\!+\!1,-m;1;\lambda)$ this limit equals to
\begin{align*}
& \sum_{k=0}^{\infty}\!\!\frac{1}{\left(k!\right)^{\!2}}\!\lim_{N\rightarrow\infty}\!\!\left[\!\left(\!N\!\!-\! m\!-\! k\!\right)\!\cdots\!\left(\!N\!\!-\! m\!-\!1\!\right)\!\!\times\!\!\left(m\!-\!(\! k\!-\!1\!)\!\right)\cdots m\!\left(\lambda^{\prime}\right)^{\! k}\!\right]\\
& =\sum_{k=0}^{\infty}\frac{1}{\left(k!\right)^{2}}\lim_{N\rightarrow\infty}\!\left[\!\left(\!1\!-\!\frac{m}{N}\!-\!\frac{k}{N}\!\right)\!\cdots\!\left(\!1\!-\!\frac{m}{N}\!-\!\frac{1}{N}\!\right)\times\right.\\ &\left.\,\,\,\,\,\,\,\,\,\,\,\,\,\,\,\,\,\,\,\,\,\,\,\,\,\,\,\,\,\,\,\,\,\,\,\,\,\,\,\,\,\,\,\,\,\,\,\,\,\,\,\,\,\,\,\,\,\,\,\,\,\,\,\,\,\,\,\,\,\,\,\,\,\,\,\!\left(\frac{m}{N}\!-\!\frac{k\!-\!1}{N}\!\right)\!\cdots\!\frac{m}{N}\!\times\left(\!\alpha\beta\!\right)^{\!k}\!\right]\\
& =\sum_{k=0}^{\infty}\frac{1}{\left(k!\right)^{2}}\left(1-x\right)^{\! k}\! x^{\! k}\!\left(\!\alpha\beta\!\right)^{\! k},
\end{align*}
because $\lim_{N\rightarrow\infty}1-\frac{\alpha}{N}=1,\,\lim_{N\rightarrow\infty}1-\frac{\beta}{N}=1$, and $\lim_{N\rightarrow\infty}\frac{i}{N}=0$ for $i=1,2,\ldots,k$. On the other hand, we know that
\[
I_{0}(z)=\sum_{k=0}^{\infty}\frac{\left(\frac{1}{2}z\right)^{2k}}{\left(k!\right)^{2}}
\]
is the zero-th order modified Bessel function. So,
\begin{align*}
& \lim_{N\rightarrow\infty}F\left(-N\!+\! m\!+\!1,-m;1;\frac{\frac{\alpha}{N}\frac{\beta}{N}}{\left(1\!-\!\frac{\alpha}{N}\right)\left(1\!-\!\frac{\beta}{N}\right)}\right)\\
& =I_{0}\left(2\sqrt{(1-x)x\alpha\beta}\right)
\end{align*}
Moreover,
\begin{align*}
& \lim_{N\rightarrow\infty}\!\left[\!(\!1\!-\!\frac{\alpha}{N}\!)^{m}(\!1\!-\!\frac{\beta}{N}\!)^{N\!-\! m}\right]=\mbox{e}^{-\alpha x-\beta(1-x)}.
\end{align*}
So,
\begin{align*}
& \lim_{N\rightarrow\infty}\,\left(N\mbox{Pr}(n_{\good}\!=\! m,s_{N}\!=\bad|s_{0}\!=\good)\right)\\
& \,\,\,\,\,\,\,\,\,\,\,\,\,\,\,\,\,\,\,\,\,\,\,\,\,\,\,\,\,=f(x,s_{N}=\bad|s_{0}=\good)\\
& \,\,\,\,\,\,\,\,\,\,\,\,\,\,\,\,\,\,\,\,\,\,\,\,\,\,\,\,\,=\alpha \mbox{e}^{-\alpha x-\beta(1-x)} I_{0}\left(2\sqrt{(1-x)x\alpha\beta}\right).
\end{align*}

However, from the first principals of probability and by \cite{Pedler-jap71} when normalizing the time interval and putting $t=1$ in the corresponding distributions for the CTMC, we know that
\begin{align*}
 & f(x,s_{N}=\bad|s_{0}=\good)+f(x,s_{N}=\good|s_{0}=\good) \\
 & =f(x|s_{0}=\good)\\
 & =\mbox{e}^{-\alpha x-\beta(1-x)}\{\delta(1-x)+\left(\frac{\alpha\beta x}{1-x}\right)^{\frac{1}{2}}I_{1}[2(\alpha\beta x(1-x))^{\frac{1}{2}}]\\
 & +\alpha I_{0}[2(\alpha\beta x(1-x))^{\frac{1}{2}}]\}.
\end{align*}
So, we can easily see that
\begin{align*}
 & f(x,s_{N}=\good|s_{0}=\good) \\
 & =e^{-\alpha x-\beta(1-x)}\{\delta(1-x)+\left(\frac{\alpha\beta x}{1-x}\right)^{\frac{1}{2}}I_{1}[2(\alpha\beta x(1-x))^{\frac{1}{2}}]\}.
\end{align*}

\subsection{A second approach to derive the conditional distributions}
In \cite{Pedler-jap71}, the derivation of of the PMF and PDF of the occupancy times for DTMC and CTMC has been done through the computation of corresponding bivariate generating function and two-dimensional Laplace transform, respectively.

First consider the DTMC. The bivariate generating function of the time spent in the good state, averaged over all initial and final states is shown to be
\[
\psi(u,x)=\left[\begin{array}{cc}
\pi_{g} & \pi_{b}\end{array}\right]\left[I-u\mathbf{G}_{\good}(x)\right]^{-1}\left[\begin{array}{c}
1\\
1
\end{array}\right]
\]
In fact, the matrix $\left[I-u\mathbf{G}_{\good}(x)\right]^{-1}$ is the bivariate generating matrix of the time spent in the good state.
For example the first entry in the matrix is
\begin{align*}
\psi_{\good\good}(u,x) & =\left[\begin{array}{cc}
1 & 0\end{array}\right]\left[I-u\mathbf{G}_{\good}(x)\right]^{-1}\left[\begin{array}{c}
1\\
0
\end{array}\right]\\
& =\frac{1-(1-\beta)u}{1-(1-\alpha)ux-(1-\beta)u+du^{2}x}.
\end{align*}
Putting $w=ux$,
\[
\Psi(u,w)=\frac{1-(1-\beta)u}{1-(1-\alpha)w-(1-\beta)u+duw},
\]
and $\mbox{Pr}(n_{\good}\!=\!m,s_{N}\!=\!\good|s_{0}\!=\!\good)$ is obtained by expanding $\Psi(u,w)$ as a power series in positive powers of $u$ and $w$. Lemma 1 in \cite{Pedler-jap71} helps to get the desired format in terms of hypergeometric functions.

For the CTMC, the matrix of two-dimensional laplace transforms of the PDF of time spent in the good state during the time interval $[0,1]$ is \begin{align*}
\left[-\left(\mathbf{Q}-\left[\begin{array}{cc}
\theta & 0\\
0 & 0
\end{array}\right]-\phi I\right)\right]^{-1} &
\end{align*}
For example the first entry in the matrix equals to
\[
\frac{1}{u}+\frac{\alpha\beta}{u(uv-\alpha\beta)},
\]
where $u=\phi+\theta+\alpha$, and $v=\phi+\beta$ and the inverse of this two-dimensional Laplace transform gives $f(x,s_{N}=\good|s_{0}=\good)$. Lemma 2 in \cite{Pedler-jap71} helps to get the desired format in terms of modified Bessel functions.

\end{document}